\documentstyle[11pt,epsfig]{article}
\begin{document}
\baselineskip 18.0pt
\def\oneskip{\vskip\baselineskip}
\def\xr#1{\parindent=0.0cm\hangindent=1cm\hangafter=1\indent#1\par}
\def\la{\raise.5ex\hbox{$<$}\kern-.8em\lower 1mm\hbox{$\sim$}}
\def\ma{\raise.5ex\hbox{$>$}\kern-.8em\lower 1mm\hbox{$\sim$}}
\def\ea{\it et al. \rm}
\def\am{$^{\prime}$\ }
\def\as{$^{\prime\prime}$\ }
\def\msol{M$_{\odot}$ }
\def\kms{$\rm km\, s^{-1} $}
\def\cm3{$\rm cm^{-3} $}
\def\Ts{$\rm T_{*} $}
\def\Vs{$\rm V_{s} $}
\def\n0{$\rm n_{0} $}
\def\B0{$\rm B_{0} $}
\def\ne{$\rm n_{e} $}
\def\Te{$\rm T_{e} $}
\def\Tgr{$\rm T_{gr} $}
\def\Tgas{$\rm T_{gas} $}
\def\Ec{$\rm E_{c} $}
\def\Fh{$\rm F_{h} $}
\def\Hb{H$\beta $}
\def\erg{$\rm erg\, cm^{-2}\, s^{-1} $}

\centerline{{\Large \bf A grid of composite models for the simulation of}}

\centerline{{\Large \bf starburst galaxy and HII region spectra}}
  
\bigskip

\bigskip

\bigskip

\centerline{ $\rm M. \, Contini^1 \,\, and \,\,\, S. M. \, Viegas^2  $}

\bigskip

\bigskip

\bigskip

$^1$ School of Physics and Astronomy, Tel-Aviv University, Ramat-Aviv, 
Tel-Aviv, 69978, Israel

$^2$ Istituto Astron\^{o}mico e Geof\'{i}sico, USP, Av. Miguel Stefano, 
4200,04301-904
S\~{a}o Paulo, Brazil

\bigskip

\bigskip

\bigskip

\bigskip

\bigskip

\bigskip

\bigskip

Running title : Grid of models for  starburst  spectra

\bigskip

\bigskip

\bigskip

\bigskip

\bigskip

\bigskip
subject headings : galaxies : starburst - shock waves -
model calculation

\newpage
\section*{Abstract}

A grid of models for starburst spectra is presented and compared with
power-law models which represent  AGNs. The models are composite, i.e., they
account for  the radiation from the starburst and for the shocks created 
by super-winds.
Some interesting fit of models to observations are also discussed.
In particular, it is found that [OIII]/\Hb ~can reach values as high as
for the AGN when the emitting clouds are close to the starburst region.
Moreover CIV lines can be high even if SiIV lines are relatively low.
It is found that for some line ratios, which have similar values
for starburst and AGN models, the width of the lines  can distinguish
between the two types.

\newpage

\section{Introduction}

One of the important issues in the interpretation of the
galactic nuclei spectra is to recognize their nature, particularly,
whether they correspond to active  nuclei or starbursts.
Modeling on a large scale may  give the key to rapidly discover  
the spectral type of observed objects.

Theoretical starburst and HII region spectra available in the literature are
calculated assuming that  
stellar radiation  ionizes and heats circumstellar clouds.
In more recent photoionization models, the ionizing radiation spectrum of
a stellar cluster is used (Cid-Fernandes et al. 1992, 
Stasinska \& Leitherer 1996).
 However, it is hardly conceivable
that a stationary regime could survive in regions around evolving stars. 
It has been suggested that, in regions with high rates of star formation in
a starburst galaxy,  an energetic, galactic scale wind - a superwind - 
may result  from the prodigious
amount of energy injected by the stars within the surrounding gas
(Chevalier \& Clegg 1986).
Among the manifestations of such a superwind (Lehnert \& Heckman 1995) 
shocklike line ratios would be of particular interest. 
As shown by the models of Leitherer \& Heckman (1995), the ratio between
the mechanical energy and the ionizing Lyman continuum, 
both injected in the ISM
by the massive star population, rapidly increases during the stellar 
cluster evolution. Thus, it is expected that the emission-line spectrum
of the ionized gas changes its characteristics, i.e.,  from a photoionized gas
spectrum to a shock-photoionized spectrum. In fact, the observed infrared
emission-line ratios of a sample of starburst galaxies (Lutz et al. 1998)
revealed a continuous sequence between two extreme cases: starburst regions
where shocks in high-velocity clouds are necessary to explain 
the observational data
and those where the shock contribution is less significant 
(Viegas, Contini \& Contini 1999).

Moreover, there have been many suggestions that starbursts may 
play an important role
in Seyfert galaxies (see, for instance, Heckman et al. 1997).
By quantifying the relative energetic importance of young stars and the
hidden Seyfert nucleus in their sample, Gonzalez-Delgado, Heckman, \& Leitherer
(2001) suggest that optical emission line diagnostics are "also potentially
useful". Particularly, they suggest that the [OIII] 5007/\Hb ~ratio
contains information about the sources of ionizing photons in the 
Seyfert 2 nuclei.

In this paper we present a grid of models for the calculation of 
starburst (STB) spectra, where the physical conditions of
the emitting gas are determined by the  effect of 
a shock front coupled to photoionization by a stellar cluster.
The SUMA code (Viegas \& Contini 1994) is adopted.
The aim of this paper is to
investigate the more relevant characteristicts of STB spectra
and compare them  with spectra calculated with a power-law (PL) 
ionizing radiation,
representing the AGN characteristic spectra
 (Contini \& Viegas 2001, hereafter referred to as CV01). 
Shock dominated (SD) models can be found in CV01.

In order to present the results, the  emission-lines 
are selected among those potentially giving
relevant information about the
source, the physical conditions of the emitting gas,
and the relative importance between the STB  and AGN characteristics.
Not only strong lines are considered,  since  sometimes weak lines
can  provide significant information. 
Therefore, a relatively large number of lines  
in the UV-optical-IR ranges (see CV01) are included in the tables.

 In \S 2, the models are presented. A comparison between STB and PL 
emission-line ratios 
appears in \S 3 and the confrontation of model results with 
observational data follows in \S 4. Concluding remarks are presented in \S 5.

\section{The models}

The SUMA code is used to obtain the physical conditions and the emission-line
spectrum of a gas cloud ionized and heated by an ionizing radiation and 
by a shock front. A plane-plarallel symmetry is assumed. Considering that 
shocks due to supernova explosions develop during the cluster evolution
and propagate outwards, it is assumed that the shock front and photoionization
act on opposite edges of the cloud. The input parameters are the 
ionizing radiation spectrum, 
the elemental abundances (H, He, C, N, O, Ne, Mg, Si, S, A, and Fe),
the shock velocity, \Vs, the preshock density, \n0, the preshock 
magnetic field, \B0, and the cloud geometrical thickness, D. For all models, 
the elemental abundances are taken from Allen(1973) and  \B0 is 
equal to  10$^{-4}$ gauss. 

The ionizing radiation spectrum is characterized by its shape and by
the value of the ionizing parameter at the photoionized edge.
Two different types of spectra are considered:

(a) A black body spectrum (BB) which could represent the spectrum of
the field stars. Thus three black body  temperatures are considered.
The lowest one (\Ts=10$^4$ K) represents the
case of rather old stars and could provide the physical 
conditions of some HII regions,whereas
\Ts=5$\times$ 10$^4$ K are characteristic of STB galaxies 
(Viegas, Contini, \& Contini 1999). 
Finally, \Ts= 10$^5$ K could represent the young stars 
and/or the "warmers" (Terlevich et al. 1992). These models are  less
realistic. However, the results can be compared to other blackbody 
photoionization  models
that have been largely used in the literature.

(b) A stellar cluster spectrum (Cid-Fernandes et al. 1992),
which is a more realistic spectrum. 
The spectrum  is  characterized by the cluster age, from 0 to 5.4 Myr. 
More evolved clusters show fewer photons above 54.4 eV, producing 
 intermediate and high ionization lines too faint compared to observations.
Notice that the mass of the stellar cluster defines its luminosity.
Thus the value adopted for the ionization parameter U is related
to the stellar cluster mass.

Before discussing the results of the models given 
 in the grid, we would like to provide
some suggestions on how to use the models to  obtain
the physical conditions of the emitting gas and  a rough
interpretation of  observed spectra, as well as 
 to select between power law and  stellar cluster spectra.

As a first approach, we suggest to  analyse a
strong line, e.g. [OIII] 5007+,
and compare models with very similar [OIII]/\Hb ~ratios.
The [OIII]/\Hb ~line ratio depends mainly  on the ionizing radiation field.
On the other hand, [OII]/\Hb ~line ratio is more sensitive to the 
velocity field. A high [OII]/\Hb ~line ratio 
indicates that shocks give a significant contribution to the emission
line spectrum. 
In addition, high velocity shocks produce a high temperature post-shock
zone which can originate high ionization emission lines.
The shock velocity is roughly indicated by  the FWHM of the 
observed line profiles and constrains the models.
Low neutral lines indicate matter-bound clouds, usually associated to 
a small geometrical thickness.
The [SII] 6717/6730 doublet ratio is a density indicator. However, 
in models accounting for shocks, the density distribution inside the
cloud results from downstream compression, which may be very different from
the preshock density \n0. Notice that clouds with different velocities and
different physical conditions may contribute to the  observed emission-line
spectrum of a single galaxy. Thus, the main difficulty is to disentangle
the contribution from single clouds. Some modeling examples can be found in our
previous papers (Contini, Prieto \& Viegas 1999a,b).

\subsection{Black-body spectra}

The results for the three black body temperatures are 
presented in Tables 1-4. 
Each table corresponds to a different value of  \Vs ~and \n0.
The ionization parameter U shows a large range (0.01-10.) 
in order to cover  most of 
the observed conditions.

In the following, a brief comparison to PL models is presented
for the different models caracterized by the (\Vs, ~ \n0) values.

First, the results corresponding to \Vs = 100 km s$^{-1}$ 
and \n0 = 100 cm$^{-3}$ (Table 1) are discussed. Unless stated
otherwise, in the following
the line results are relative to \Hb.

High [OIII]5007+\Hb ~($\geq 10$), which are   characteristic 
of the NLR spectra in Seyfert 2 galaxies,
could be produced also in the surroundings
of a hot star with \Ts=10$^5$ K . 
Compared with a PL model (CV01, model 13)
which also shows a high [OIII]5007+/\Hb, the BB models show fainter 
low ionization line ratios. 

High [OIII]/\Hb ~($\sim$ 16) can also be found for PL
models with log \Fh ~between  $\sim$ 11.5 - 11.6 (CV01 models 16 and 17). 
In this case the gas is highly ionized, the low ionization
zone is reduced, so the 
low ionization lines are negligible, while coronal lines in the IR
are  strong ([SiIX], [MgVIII], etc). Notice that BB models with
\Ts=$10^5$ K and U=1 can provide [OIV]/\Hb ~ stronger than PL models.
This BB model also show high HeII 1640 and CIV 1550. 

 LINER spectra are characterized by [OIII]/\Hb ~ of about 5.
 The choice of the model depends on the Fe coronal lines. If
[FeX] ~and [FeXI] ~are strong, PL models with 
high \Fh ~are required. However, if Fe coronal lines are negligible
the critical line  is [SIII] 9532+.
See for instance, model 12 from CV01, which shows negligeble
Fe coronal lines, but [SIII] stronger by a factor of $\sim$ 10,
when compared to the BB results.

The results calculated with \Vs=200 \kms ~and \n0=200 \cm3 appear in Table 2.
 [OIII]/\Hb ~ $\sim$ 4-10  results from BB models (\Ts=5 $10^4$ K 
and \Ts =10$^5$ K). They correspond to PL models with 
log \Fh = 11.48-11.78 (CV01, models 36 and 37).
Higher  [OIII]/\Hb ~($\sim$ 22-40) result from  
\Ts=10$^5$ K  corresponding 
to PL models with log \Fh=12-12.48 (CV01, models 38 and 39). 
However, UV line  are definitively stronger in the PL case,
as also the  \Hb ~ absolute flux. Infrared coronal lines are also stronger  
in the PL case, while neutral and singly ionized lines in the infrared 
([NeII], [FeII], [CI], [CII], etc)
are definitively higher for BB models.
[OIII]/[OII] is less than 1 for \Ts=$10^4$ K.

The results of models calculated with \Vs=300 \kms ~and \n0=300 
\cm3 are given in Table 3.
\Hb ~ absolute values are  strong for both  PL and BB models, therefore, 
the line ratios are relatively low.
SD models are characterized by higher line ratios because \Hb ~ is very low.

Regarding the results of models calculated for \Vs=500 \kms 
~and \n0=300 \cm3 (Table 4),
the spectra resulting from  BB models, by varying \Ts ~and U,
 are very similar  because the effect of the shock and, particularly, 
of diffuse radiation  prevails.
\Hb ~ absolute flux is very high  ( $>$ 600 \erg).
The models of Table 4 are all characterized by strong HeII  both in the UV 
and in the optical ranges. In  a similar PL case (CV01, 
model 66, log \Fh = 12)  [OII]/\Hb ~ratio is 
slightly higher and CIV lines are weaker.
In the IR, [OIV] and [NeIII]15.5 lines are strong in both cases.
High ionization lines, relative to \Hb, ~are  faint not only because 
\Hb ~is strong, but also due to compression which reduces the volume
of the zone where they can be produced.

\subsection{Stellar cluster models}

The emission-line results are presented in Tables 5-10.
The models are calculated adopting D=10$^{19}$ cm in Tables 6, 8, 10.
Matter-bound models calculated with D = 10$^{17}$ cm are given for \Vs=100 and 
300 \kms (Tables 5, 7) and with D = 10$^{18}$ cm for \Vs=500 \kms ~(Table 9).
A higher D is adopted for \Vs = 500 \kms
~because a large zone of highly ionized gas is created downstream by high
shock velocities (see CV01). The matter-bound models are presented 
for comparison and for helping modeling. However, the narrower the cloud
the smaller the low-ionization zone located at the cloud central region.
Thus, for more realistic modeling many different D should be considered. 
In fact, fragmentation
produced by the presence of shocks (through turbulence and
instability effects)
can reduce the geometrical thickness of the clouds by different amounts.
Thus, when modeling specific objects the contribution of 
clouds with different geometrical thickness  should be taken into account. 
Notice that matter-bound and radiation-bound models
depend on U and \Vs, as well as on geometrical thickness D.
Models corresponding to D=10$^{17}$ cm are generally matter-bound, 
as indicated, for instance, by the low theoretical values of 
 the [OI]/\Hb ~and [NI]/\Hb ~line ratios.

A general look at the tables indicates that for \Vs=100 \kms and \n0=100 \cm3
UV lines (1033-1892 \AA) are stronger for t $\geq$ 3.3 Myr. 
In the optical range
(3426-7892 \AA) some lines
(e.g. [OII]3727, [OIII]5007, HeI 5876, [NII]6548+, and [SII] 6716,6730)
can also be relatively strong at a lower age, however, they drastically
increase for t $\geq$ 3.3 Myr and relatively low U. In the near-infrared range 
(9532 \AA - 3.9 $\mu$m)
the lines, which are mostly emitted from high ionization states, 
are observable only for STB of a higher age.  In the infrared
 range (5.5 - 157.7 $\mu$m),
high ionization level lines are strong for t $\geq$3.3 My, 
intermediate ionization
lines (e.g., [SIV]) are relatively strong also at lower ages with high U.
Low ionization lines are stronger the lower U.
For \Vs=300 \kms ~and \Vs=500 \kms ~the effect of the shock is strong,
and the line ratios  change sensibly  only for t $\geq$ 3.3 Myr. 

\section{Confrontation of STB versus PL model predictions}

It is our goal to give a general discussion of the
models in order  to understand their behavior
regarding the characteristics of the ionizing radiation and of the shock. 
In the following, only the comparison between STB and PL radiation-bound
models is presented. The notation used in the figures is:
starburst  models corresponding to  different ages are 
represented by dotted 
(t = 0 Myr, SBO models), short-dashed (t = 2.5 Myr, SB2 models)  long-dashed 
(t = 3.3 Myr, SB3 models), short-dash-dotted (t = 4.5 Myr, SB4 models), and
long-dash-dotted (t = 5.4 Myr, SB5 models) lines.
Power-law models are represented by a solid line.
Shock-dominated models are represented by filled circles
(in diagrams corresponding to a single \Vs) and  by a 
long-dash-short-dash line.
When models corresponding to different values of the shock velocity are
shown in the same figure, heavier lines refer to  the highest \Vs.
For sake of clarity, only results corresponding to models 
with  D = 10$^{19}$ are plotted (Tables 6, 8, and 10).

Recall that this paper is aimed to present  a grid of results of STB models,
therefore, only a few significant cases are presented in the diagrams.
Particularly, in the UV range  the Ly$\alpha$/\Hb
~line ratio shows very high values which could explain 
uncommon  observational results. 
Other permitted lines, as e.g.  CIV, NV and SiIV, 
are strongly connected with star
formation and are therefore discussed.
In the optical-IR range we have chosen lines of the 
same element which are generally
observed  from different levels, as e.g. [NeIII] 3869+, [NeV] 3426, 
[NeIII] 15.55 and [NeV] 24.3. The
[NeIII] 3869+/\Hb ~behavior is similar to that of [OIII] 5007+/\Hb.
The calculated O lines ([OIII] 5007+, [OII] 3727, and [OI] 6300)
 will be  compared   with observational data. 

\newpage

\subsection{UV lines}

\bigskip

\centerline{\it Ly$\alpha$ and HeII 1640}

\bigskip

Ly$\alpha$/\Hb ~versus HeII 1640/\Hb ~is given in Fig. 1 for SB3, SB4, and SB5 
models and different \Vs.
For SB1 and SB2 the calculated HeII lines are low   for \Vs=100
and 300 \kms ,
and they are constant for \Vs=500 \kms ~and different U, reducing to
a point in the diagram (see Tables 6, 8 and 10).

The high jump of Ly$\alpha$ /\Hb ~for U $\leq$ 0.1 (SB3) and \Vs=100 \kms, 
which appears also  for U=0.01 (SB4 and SB5), 
is investigated in Figs. 2. Here, the physical conditions and the 
H fractional abundance
are shown in two cases: for U=0.1 (top diagram) and U=1 (bottom diagram),
corresponding  to Ly$\alpha$/\Hb =106. and 24.8, respectively, and
H$\alpha$/\Hb = 4. and 2.77, respectively. 
For U = 0.1, the gas is heated  to 10$^4$ K in a region where
hydrogen has a significant neutral fraction, consequently 
(Ferland \& Osterbrock 1985), the intensities of H$\alpha$ 
 and especially of Ly$\alpha$ are significantly enhanced, 
although the intensity of \Hb ~remains near its pure recombination 
value (because 1-4 collisions are much less 
frequent than 1-3 or 1-2 collisions).
For U = 1 (bottom diagram) the ionization parameter is 
high enough  to fully ionize hydrogen
throughout the  cloud and the value of Ly$\alpha$/\Hb ~is lower.

The He II 1640/4686 versus He II 1640/\Hb ~ratios are shown in Fig. 3.
Notice the gap between SD models and RD models,  for both models with PL
and STB ionizing spectra.

\bigskip

\centerline{\it NV 1240, SiIV 1397, and CIV 1549}

\bigskip

Starburst galaxies have been investigated through the massive star wind
properties by Robert et al. (1993).
 Unambiguous spectral features
from massive stars,  which display characteristic profiles of SiIV 1397 and
CIV 1550),  are most  easily found in the UV (Weedman et al 1981).
Robert et al. consider a burst model with a finite duration of the star
formation process. The emission-line SiIV 1397 is due to massive stars
 and is strong only for a short time, around 10$^{6.5}$
yr, in the case of the instantaneous burst.
Such a strong emission 
 is not seen in the continuous burst models since main-sequence stars, 
which are continuously being born, dilute the SiIV contribution 
coming from the more massive evolved stars.
A strong SiIV  emission is rarely, if ever, seen 
in starburst galaxies.

Our approach, is, however, different, because we consider the lines
emitted by clouds lighted by a STB  (e.g. NV, SiIV, CIV, etc)
to test for either a STB component or a AGN one in the UV spectra.

In Fig. 4, the results of PL models and STB models are
shown in the left and right panels, respectively. PL models
show an increase of NV/\Hb ~proportionally to  
CIV/\Hb. Models  with high velocities 
($\geq$ 300 \kms)
give low line ratios. This is explained by the strong compression donwstream
which reduces the emission region of lines from the IV and V levels,
whereas an extended region of \Hb ~emission is supported by a large \Fh.
The trend is different for STB models (top right diagram).
The line ratios increase for higher velocities. Moreover,
they give  strong CIV/\Hb  ~even for low NV/\Hb.

A maximum value of the order of unity is reached by SiIV/\Hb ~lines
(bottom diagrams). Regarding the PL models, the maximum values is reached
for PL models with  \Vs=100 \kms ~and  log \Fh=11.48. On the 
other hand,  the maximum value can be found in several
STB models and there is not a univocal relationship between CIV and SiIV
line intensities.

Notice that SiIV/\Hb ~calculated line ratios are very low. Let us recall 
that, the  observed SiIV lines 
could be  blended with those of the OIV 1400 multiplet due to wavelength
proximity. 
Therefore, they must be used with caution.

\newpage

\subsection{Optical-IR lines}

\bigskip

\centerline{\it [NeIII]3869,[NeV]3426, [NeIII]15.55, and [NeV]24.3}

\bigskip

Usually neon is less depleted than other elements because it 
is not easily locked up into dust grains  owing to its electronic
configuration.

We have chosen lines from the same ions ( Ne$^{+2}$ or Ne$^{+4}$) 
 but in different
ranges (optical and IR) to search for similarities and differences
between  STB and PL models.  The results are shown in Figs. 5
for \Vs=100, 300, and 500 \kms.

The first three panels (Figs. 5a, b, and c) show that when 
the shock velocity is low  (\Vs=100 \kms) and photoionization
prevails on shock
effects,the results from  PL models, with  F$>$ 10$^{11}$ 
photons cm$^{-2}$ s$^{-1}$ eV$^{-1}$,
are similar to those  from SB3, SB4, and SB5 models with  
U $>$ 0.1. A younger stellar cluster  (SB0 models)
give similar results to PL ionizing source  in a small range of \Fh 
~(log \Fh ~between 8 and 9.48). 
On the other hand, for higher \Vs, SB3, SB4, and 
SB5 models  show  higher [NeIII]3869/\Hb  ~ratios  if U $>$  0.1.
For \Vs=500 \kms all the values for SB0 and SB2 are concentrated
in one point, with [NeIII] 3869/\Hb ~about 11 and [NeIII]/\Hb ~about 2.

Fig. 6  shows that high [NeV]/[NeIII] ratios  both in the
optical and IR ranges favors the starburst models, although
 PL models with \Vs=100 \kms and a high \Fh ~can also
give a high [NeV]/[NeIII]. 
Notice that, in Fig. 6, only SB3, SB4, and SB5 models are shown. 
Because young stellar clusters have a deficit in high energy photons,  
SB0 and SB2 provide  very   low values of [NeV]/\Hb, not seen in HII regions.
So, high [NeV]/[NeIII] line ratios showing 
broad profiles of the single lines refer to starbursts.

\section{Comparison with  observations}

When observational data are available, we prefer to compare
our  theoretical lines with those observed in different locations
of a given galaxy, rather than  data from several galaxies.

The phenomena STB and AGN can coexist in  one galaxy. 
Thus, when adopting a single average model 
as representative of a galaxy, as generally used
in diagnostic diagrams
(see Veilleux \& Osterbrock 1993) a large  part of information is lost.
Therefore, we have chosen NGC 7130 that was observed in different regions
(Shields \& Filippenko 1990, Radovich et al. 1997, etc) for modeling
the most  observed oxygen lines. 

Observational data for the sample of edge-on STB galaxies
by Lehnert \& Heckman (1995)are available for  different position
within each galaxy. The [SII], [NII], and [OI] lines are considered.
The distribution and scattering of the data in the nuclear, near-nuclear, and 
off-nuclear regions indicate the gas conditions.

II Zw 40 is one of the STB galaxies observed by Doherty et al (1995).
 Different values of the He I 5876/\Hb ~ratios, corresponding to different 
observations, are reported in  Doherty et al. (Table 4 there).

Finally, [SIII] 9532 emission is chosen as another example 
in order to illustrate that 
the discussion about the range of the [SIII]/H$\alpha$
line ratios (Kennicutt \& Pogge 1990) can be settled by comparing the data
with model results.

\bigskip

\centerline{\it [OI]/\Hb, [OII]/\Hb, [OIII]/\Hb}

\bigskip

The O lines  [OI]6300+, [OII]3727, and [OIII]5007+ 
are generally observed in AGN and starburst galaxies, and
provide an indicator of the shape of the ionizing radiation 
spectrum. The results for these lines appear in Fig. 7a and b
and are compared to the observational data from the Seyfert 2 
galaxy NGC 7130 (Shields \& Filippenko 1990). 
The spectra were taken in different regions at P.A.=11.5$^o$.

It is well known that this Seyfert galaxy contains starbursts in the
circumnuclear regions.
Our aim is to show that the models can explain the observations.
As this work focus on the models and not on a particular
investigation of special objects,  we display in the figures not only
the  diagram area  limited by the data, but we include a larger area  in order
to  show many model results. 

The results are shown for shock velocities of 100, 300, and 
500 \kms (top, middle, and bottom diagrams, repectively).
Shock-dominated model results, corresponding to \Fh = 0,
are indicated as filled circles.

For low velocity, PL models are similar to SB3, SB4, 
and SB5 models (Fig. 7a)
For higher \Vs, ~the results are strongly affected by compression
and by the   physical conditions downstream. For \Vs=300  and 500 \kms, 
[OI]/\Hb  ~as high as 1 is  obtained only by PL models with a
low \Fh. The  STB models  give smaller values. Particularly, SB3 models show
higher [OIII]/\Hb ~to [OI]/\Hb ~ratios relative to PL models.

A better fit is obtained for low velocities  and  low ionization parameters,
and a young stellar cluster (t = 0.0 to 2.5 Myr).
For a few data near the center of the galaxy,  PL models give a better fit.

The same  general considerations  are valid for the 
[OIII]/\Hb ~versus [OII]/\Hb  ~ratios, shown in in Fig. 7b.
Notice, however, that models calculated with low ionization parameter
(U = 0.01) give a better fit of  the [OII]/\Hb
~than of the [OI]/\Hb ~ratios. Models with higher U better reproduce the 
neutral lines.  Therefore, we prefer to explain this
discrepancy between the [OI]/\Hb ~and [OII]/\Hb  ~fits
by adopting matter-bound models (Contini et al 2001).

\bigskip

\centerline{\it [NII]/\Hb ~and [SII]/\Hb}

\bigskip

The NII and SII lines are also easily observed.
In Figs. 8a,b model results are compared with the observation data of
Lehnert \& Heckman (1995) for ionized gas in the halos
of edge-on starburst galaxies. The nuclear, near-nuclear, and off-nuclear
data appear in the top, middle, and bottom diagrams, respectively.
The large scattering, particularly in the near- and off-nuclear
regions indicate that many different conditions characterize
the different galaxies of the sample. 
Observations corresponding to the minor and major axis are represented 
respectively by  filled and open symbols.
For near- and off- nuclear data the observations come from two
positions on each  axis (major and minor) of the galaxy. 
The data have been  multiplied by a factor of 3 in order to obtain
the line intensities relative to  \Hb ~instead of H$\alpha$.

The diagram  [SII]/\Hb ~versus [NII]/\Hb ~is shown in Fig. 8a.
 It can be noticed 
that, in all of three diagrams, the data are 
inside a region covered by  STB models.

By reducing S/H by a factor of 2-3, the fit with
models calculated for a t = 2.5 Myr and a \Vs ~of 100 \kms ~improves.

However,  PL models with \Vs=300 \kms 
~and a rather low \Fh,  as well as shock-dominated models 
with \Vs $\leq$ 300 \kms ~could also provide a good fit to the data 
(PL models calculated with \Vs = 100 and 500 \kms ~do not appear
in the diagrams for sake of clarity; moreover, they are out of the observed range).
Once more, this indicates that single models can 
hardly be used and that [SII] and [NII] lines are not a good indicator
of the galaxy type, whether a STB or a AGN. 
Observational data correspond to emission coming from regions
with different physical conditions not easily fitted by a single model. 

On the other hand, concerning the diagram [OI]/\Hb ~versus [NII]/\Hb 
~(Fig. 8b), a  better fit to the data is obtained with STB models, 
strengthening  the hypothesis that the less precise fit  
in Fig. 8a could be due to the cosmic S/H abundance adopted in the model,
as pointed out above.
Notice that upper limits of [OI]/\Hb ~are also included in Fig. 8b diagrams.

\bigskip

\centerline{\it HeI 5876/\Hb ~and HeI 4471/\Hb}

\bigskip

 A heterogeneous
sample of starburst galaxies is used to analyse the He lines in Fig. 9
(PL models are not included in the figure for sake of
clarity, because they concide with STB models).
The data come from Doherty, Puxley, Lumsden, \& Doyon (1995).
STB models calculated for an age $\geq$  3.3 Myr, U$\leq$ 1,
and  \Vs ~between 100 and 300 \kms ~fit  the data corresponding to
the highest He I 5876/\Hb ~ratios. Lower 
 ratios are not fitted  by the present models.
Particularly, shock-dominated models and STB models corresponding 
to a young stellar cluster  are out of the observed range. Notice,
however, that the HeI lines can be affected by the geometrical
thickness of the emitting cloud. A narrower cloud could have the
He$^+$ emitting zone reduced, producing  fainter HeI lines.

\bigskip

\centerline{\it [SIII] 9532 emission}

\bigskip

Observations of [SIII]/(H$\alpha$+[NII]) line ratios for HII regions in
infrared-luminous galaxies (NGC 520, NGC 660, NGC 1068, NGC 2146, and M82)
are given by Kennicutt \& Pogge (1990) to
 critically examine these ratios as quantitative extinction 
tracer in HII regions.
The comparison of the data (corresponding to [SIII]/(H$\alpha$+[NII])
between 0.2 and 1.18) 
with the results of some models in Tables 5-16 gives a good agreement.

\section{Concluding remarks}

By presenting a grid of composite models for STB galaxies
we have shown that the ionization parameter and the
shock velocity play an important role.
We have calculated the emission-line spectra from clouds in the neighborhood
of a STB within a galaxy, considering both the radiation and the 
effect of a super-wind. 

Particularly, we have found that only STB and SD models can give a very
high Ly$\alpha$/\Hb ~ratio. The result for the STB has been explained
by the very special profile of H$^+$ fractional abundance throughout a cloud.

Another important result concerns the CIV lines. It is found 
that CIV/\Hb ~can be high even if NV/\Hb ~and, particularly, 
SiIV/\Hb ~are relatively low.
 This happens in clouds very close to the
STB, for an age of t $\geq$ 3.3 Myr. 
High CIV and  NV lines correspond
 to \Vs ~of  about 500 \kms ~in the STB, and to \Vs ~of about 100 \kms 
~in the AGN case.
 {\it The FWHM of the line profile permits to 
distinguish between the two types}.
For SiIV,  similar low  values result
also from PL models, which would correspond to AGN emission-line spectra. 
The SiIV emission line is usually faint since, 
due to the low SiIV ionization potential, SiV is the dominant ion 
and the SiIV emitting volume is small.

Very high [NeV]/[NeIII] in both the optical and IR domains are always
indicative of a STB, if the FWHM of the line profile is $>$ 100 \kms.

The results of a large grid of models show that
[OIII]/\Hb ~can be high ($>$ 10) also for STB, although it is  generally
suggested that [OIII]/\Hb $\leq$ 1 (Gonzalez Delgado et al 2001).
This implies U $>$ 0.01 and leads to  [OI]/\Hb $\leq$ 1 for \Vs = 100 \kms.
For higher \Vs, ~[OI]/\Hb ~is very low ($\leq$ 0.01). Because 
of the low critical density for collisional 
deexcitation (about 3000 \cm3), the [O II] optical line
is less adapted to modelling in a general way.

Regarding the Lehnert \& Heckman sample of starburst galaxies, 
it seems that  sulphur should
be underabundant to give a better fit  to the data (see Fig. 8a). 
On the other hand, Fig. 8b shows
that the data are mostly explained  by STB  models, particularly in the
nuclear region. 
Although not shown in the figure, PL models  corresponding to 
\Vs=100 \kms  ~overpredict the [OI] lines, while the those with 
 \Vs=300 \kms ~and  \Vs=500 \kms ~poorly fit the observational trend.

For low ionizing radiation intensities 
the models are strongly shock dominated, confirming the 
important role of shocks in
modeling the low ionization line ratios.

Regarding HeI 5876/\Hb, model results indicate that  HeI/\Hb 
$\leq$ 0.08 for II Zw 40 are not well reproduced by the models
presented here, but models with narrower emitting clouds
may provide a better fit to the observations.

In a forthcoming paper (Contini, Viegas \& Contini, in preparation)
we use the models of the present grid to analyse
the emission-line spectra of starburst galaxies, comparing
young and old starburst dominated objects, as was done 
in a previous paper (Viegas et al. 1999).

Concluding, we hope that the grid presented in this work will provide more 
understanding about the physical conditions in STB galaxies.

We strongly recommend to consider single observed spectra in each 
 region of a galaxy during the modeling process.  
The final result should then be obtained by a weighted average 
in order to have a more reliable picture of the conditions prevailing
in the emittion region.
It is well known that many different conditions coexist.
Therefore, modeling  a single global spectrum from a galaxy can 
give only a first, rough hint  of the  real conditions.

\bigskip

Acknowledgements: The authors are indebted to an anonymous referee
for helpful comments. This paper is partically supported by the 
Brazilian financial agencies CNPq (304077/77-1), PRONEX/Finep
(41.96.0908.00) and FAPESP(00/06695-0).

\newpage

{\bf References}

\bigskip

%{\magnification = 1200}
%{\pageno = 14}
\vsize=26 true cm
\hsize=15 true cm
\baselineskip=18 pt
%
% def. para linhas de referencias (Bia)
\def\ref {\par \noindent \parshape=6 0cm 12.5cm 
0.5cm 12.5cm 0.5cm 12.5cm 0.5cm 12.5cm 0.5cm 12.5cm 0.5cm 12.5cm}

\ref Allen, C.W. 1973, Astrophysical Quantities (London: Athlon)
\ref Chevalier, R.A. \& Clegg, A.W.  1985, Nature, 317,44
\ref Cid-Fernandes, R., Dottori, H.A., Gruenwald, R.B., \& Viegas, S.M. 1992, 
MNRAS, 255, 165
\ref Contini, M., Prieto, M.A., \& Viegas, S.M. 1999a, ApJ 492, 511
\ref Contini, M., Prieto, M.A., \& Viegas, S.M. 1999b, ApJ 505, 621
\ref Contini, M., Radovich, M., Rafanelli, P., \& Richter, G. 2001, submitted
\ref Contini, M. \& Viegas, S.M. 2001, ApJS, 132, 211 (CV01)
\ref Doherty, R.M., Puxley, P.J., Lumsden, S.L., \& Doyon, R. 1995, 
MNRAS, 277, 577
\ref Ferland, J.G. \& Osterbrock, D.E. 1987, ApJ, 289, 105
\ref Gonzales Delgado, R.M., Heckman, T.M., \& Leitherer, C. 2001, 
ApJ, 546, 845
\ref Heckman, T.M. et al 1997  1997, ApJ, 482, 114
\ref Kennicutt, R. C.  \& Pogge, R.W. 1990, AJ, 99 61
\ref Lehnert, M.D. \& Heckman, T.M.  1995, ApJS, 97, 89
\ref Leitherer, C. \& Heckman, T.M. 1995, ApJS 96, 9
\ref Lutz, D., Kunze, D., Spoon, H.W.W. \& Thornley, M.D. 1998, A\&A 333,L75
\ref Radovich, M., Rafanelli, P., Birkle, K., \& Richter, G. 1997, 
Astron. Nachr., 318, 229
\ref Robert, C., Leitherer, C., \& Heckman, T.M. 1993, ApJ, 418, 749
\ref Shields, J.C. \& Filippenko, A.V. 1990, AJ, 100, 1034
\ref Stasinska, G. \& Leitherer, C. 1996, ApJ, 107, 661
\ref Viegas, S.M. \& Contini, M. 1994, ApJ, 428, 113
\ref Viegas, S.M., Contini, M., \& Contini, T. 1999, A\&A, 347, 112 
\ref Terlevich, R., Tenorio-Tagle, G., Franco, J., Melnick, J. 1992, MNRAS,
255, 713
\ref Weedman, D.W., Feldman, F.R., Balzano, V.A., Ramsey, L.W., Sramek, 
R.A., \& Wu, C.-C.  1981, ApJ, 248, 105

\newpage

\centerline{\bf Figure Captions}

\noindent
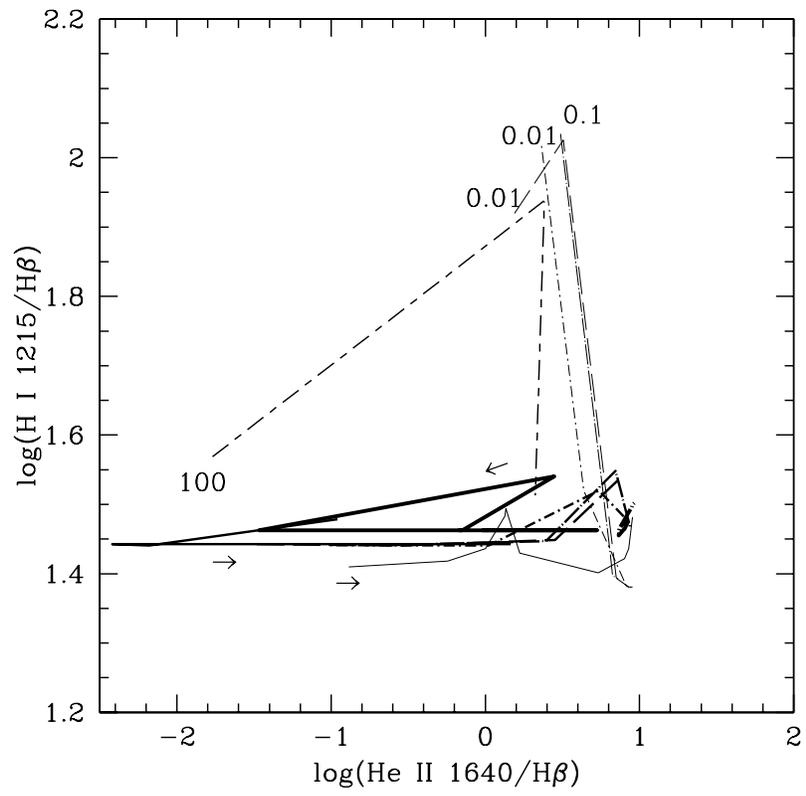
{\bf Fig. 1}

Ly$\alpha$/\Hb ~versus HeII 1640/\Hb. 

Dotted, short-dashed, long-dashed, short-dash dotted, long-dash 
dotted lines refer to STB
 models  (see text) and are labeled by U.
Solid lines refer to PL models. The arrow indicates the direction of
increasing \Fh. Concerning models with different velocities,
the notation is the following:
\Vs=100 \kms : thin lines, 300 \kms : medium lines, and 500 \kms :
thick lines.
Long-dash-short-dash lines refer to shock-dominated (SD) models, 
labelled by the 
lowest \Vs. The results correspond to models with D = 10$^{19}$ cm.

\bigskip

\noindent
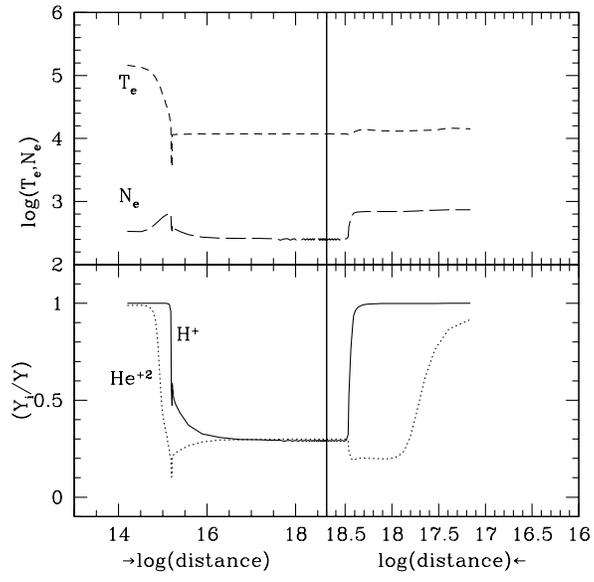
{\bf Fig. 2}

The distribution of the electron temperature, of the electron density,
and of the fractional abundance of H$^+$ and O$^{+2}$ throughout a cloud
with D=10$^{19}$ cm, \Vs=100 \kms, and \n0=100 \cm3.
Top diagram : U=0.1, bottom diagram : U=1. The diagrams are symmetrically
divided in two halves to better compare the conditions in the shock
dominated and radiation dominated edges. The shock front is on the left.

\bigskip

\noindent
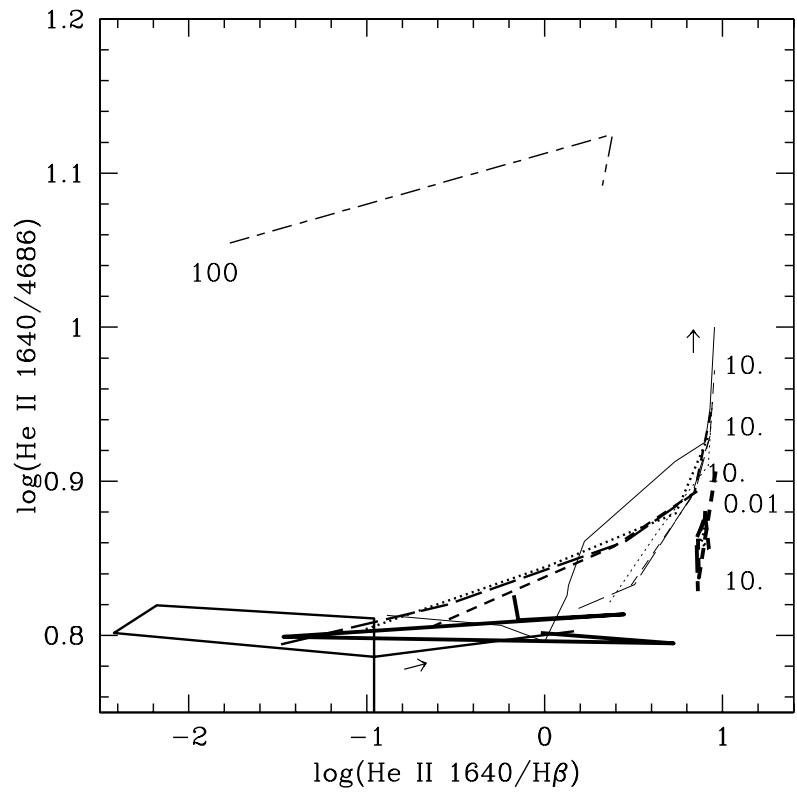
{\bf Fig. 3}

HeII 1640/ HeII4686 versus HeII 1640/\Hb.
Notation as in Fig. 1.

\bigskip

\noindent
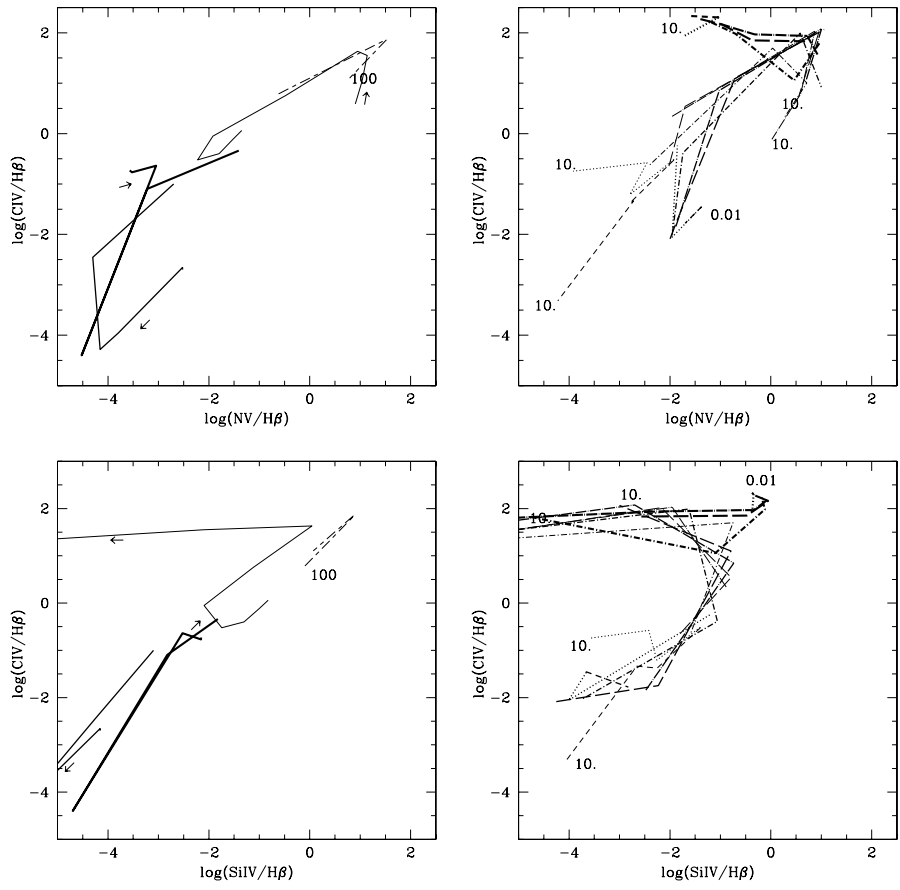
{\bf Fig. 4}

Top diagrams :  CIV 1550/\Hb ~versus NV 1240/\Hb;
left : PL models, right : STB models. 
Bottom diagrams :
CIV 1550/\Hb ~versus SiIV 1397/\Hb;
left : PL models, right : STB models.
Notation as for Fig. 1. See also the text.

\bigskip

\noindent
{\bf Fig. 5}

[NeIII] 15.55/\Hb ~versus [NeIII] 3869/\Hb ~for three different \Vs:
a) 100 \kms, b) 300 \kms, c) 500 \kms. Notations as for Fig. 1,
except for  SD models which are represented by a filled circle.

\bigskip

\noindent
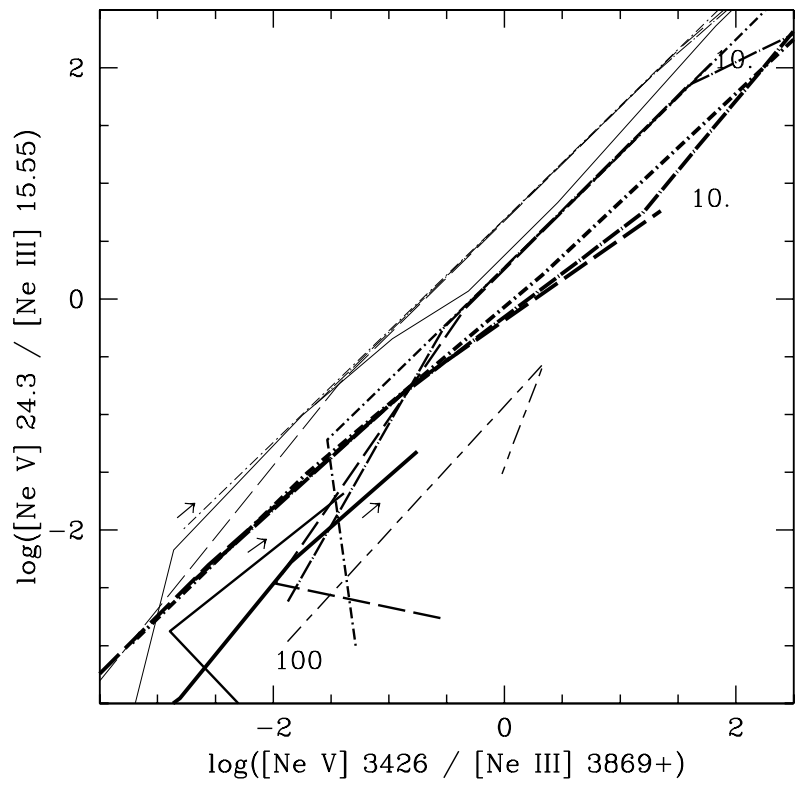
{\bf Fig. 6}

[NeV]24.3/[NeIII] 15.55 versus [NeV] 3426/[NeIII]3869.
Notation as in  Fig. 1.

\bigskip

\noindent
{\bf Fig. 7}

a) [OIII] 5007+/\Hb ~versus [OI] 6300+/\Hb ~for \Vs=100 \kms (top diagram),
300 \kms (middle diagram), and 500 \kms (bottom diagram). 
Notation as in Fig. 4.
Observational data are indicated by filled triangles. 

b) [OIII] 5007+/\Hb ~versus [OII] 3727/\Hb ~for \Vs=100 \kms (top diagram),
300 \kms (middle diagram), and 500 \kms (bottom diagram). Notation as in
Fig. 6a.

\bigskip

\noindent
{\bf Fig. 8}

a) [SII] 6717+6730/\Hb ~versus {NII] 6548+6584/\Hb. Models  are compared
with data in the nuclear (top diagram), near-nuclear (middle diagram),
and off-nuclear (bottom diagram). Notation as in Fig. 1.
Filled triangles refer to observations along
 the minor axis and   empty triangles
along  the major axis. Squares represent the data on the same axis but on
the other side of the nucleus.

b)  [OI] 6300+6363/\Hb ~versus [NII] 6548+6584/\Hb, using the
same notation as in Fig. 7a. 

\bigskip

\noindent
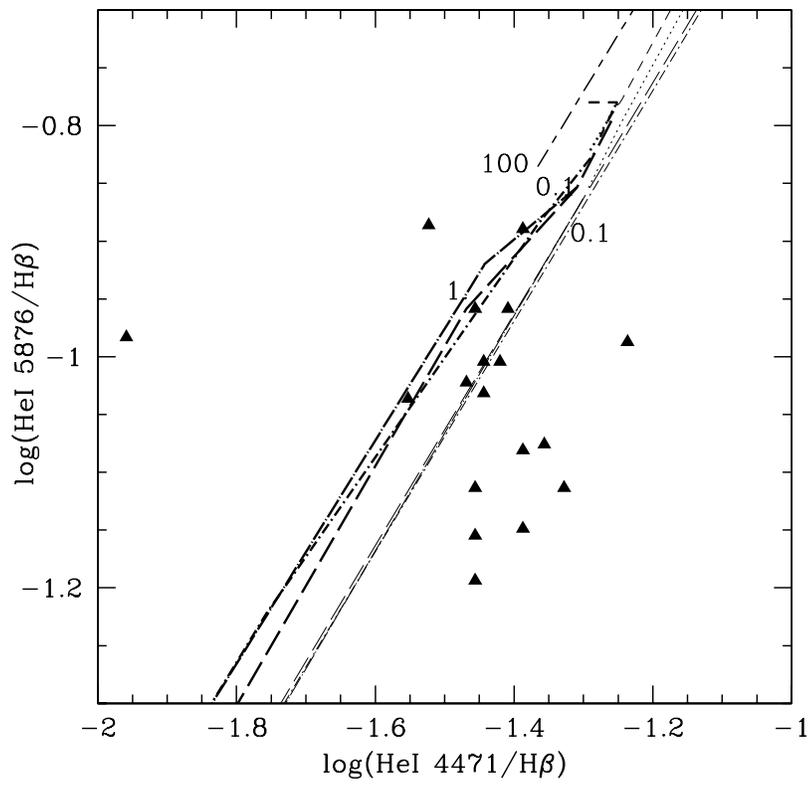
{\bf Fig. 9}

HeI 5876/\Hb ~versus HeI 4471/\Hb ~for starburst galaxies. 
Notations as in Fig. 1. 
Observational data correspond to filled triangles.

\end{document}